\documentclass[
    amsmath,
    amssymb,
    superscriptaddress,
    aps,
    prx,
    reprint,
    floatfix,
]{revtex4-2}

\usepackage{graphicx}
\usepackage{dcolumn}
\usepackage{hyperref}
\usepackage{glossaries}
\usepackage[version=4]{mhchem}
\usepackage[per-mode=symbol]{siunitx}
\usepackage{xcolor}
\usepackage{booktabs}

\hypersetup{
    unicode,
    colorlinks=true,
    urlcolor=black,
    linkcolor=black,
    citecolor=black
}

\renewcommand{\vec}[1]{\ensuremath\boldsymbol{#1}}

\newacronym{afe}{AFE}{antiferroelectric}
\newacronym{ddo}{DDO}{dynamically disordered off-centering}
\newacronym{dft}{DFT}{density-functional theory}
\newacronym{fe}{FE}{ferroelectric}
\newacronym{md}{MD}{molecular dynamics}
\newacronym{mlip}{MLIP}{machine-learned interatomic potential}
\newacronym{qnep}{qNEP}{charge-aware neuroevolution potential}

\newcommand{\gpumd}{\textsc{gpumd}}
\newcommand{\ase}{\textsc{ASE}}
\newcommand{\calorine}{\textsc{Calorine}}
\newcommand{\ovito}{\textsc{ovito}}

\makeatletter
\let\oldtheequation\theequation
\renewcommand\tagform@[1]{\maketag@@@{\ignorespaces#1\unskip\@@italiccorr}}
\renewcommand\theequation{(\oldtheequation)}
\makeatother

\DeclareSIUnit\angstrom{\text{Å}}
\DeclareSIUnit\atom{\text{atom}}
\DeclareSIUnit\elementarycharge{\ensuremath{e}}
\DeclareSIUnit\bohr{\ensuremath{a_0}}

\newcommand{\hmn}[1]{
  \ensuremath{\begingroup\setupHMN #1\endgroup}%
}

\newcommand{\setupHMN}{%
  \doHMN{-}{\HMNoverline}%
  \doHMN{*}{\HMNminverse}%
  \doHMN{i}{\infty}
}

\newcommand{\doHMN}[2]{%
  \begingroup\lccode`~=`#1
  \lowercase{\endgroup\let~}#2%
  \mathcode`#1="8000
}

\newcommand{\HMNminverse}[1]{\frac{#1}{m}}
\newcommand{\HMNoverline}[1]{\mkern1mu\overline{\mkern-1mu#1\mkern-1mu}\mkern1mu}

\begin{document}

\title{Interplay of B-Site Off-Centering and Molecular Orientations in the Mixed Hybrid Perovskite \ce{MAGe_{1-x}Sn_xI3}}

\newcommand{\addchalmersphy}{Department of Physics and Astronomy, Chalmers University of Technology, SE-412 96 Gothenburg, Sweden}
\newcommand{\addchalmerschem}{Department of Chemistry and Chemical Engineering, Chalmers University of Technology, SE-412 96 Gothenburg, Sweden}

\author{Erik Fransson}
\affiliation{\addchalmersphy}

\author{Apinya Ngoipala}
\affiliation{\addchalmersphy}

\author{Oskar \"Ojstedt}
\affiliation{\addchalmerschem}

\author{Maths Karlsson}
\affiliation{\addchalmerschem}

\author{Paul Erhart}
\affiliation{\addchalmersphy}

\author{Julia Wiktor}
\email{julia.wiktor@chalmers.se}
\affiliation{\addchalmersphy}

\date{\today}

\begin{abstract}
B-site mixing is a common strategy for tuning properties of halide perovskites.
In the lead-free system \ce{MAGe_{1-x}Sn_xI3}, it brings tilting and off-centering into competition.
Using large-scale molecular dynamics driven by a machine-learned interatomic potential, we map the structural behavior across the full composition range.
\ce{MAGeI3} exhibits strong polar B-site off-centering that remains nearly constant up to the cubic transition, together with methylammonium (MA) orientational order that weakens progressively on heating.
By contrast, \ce{MASnI3} combines octahedral tilting with weaker, predominantly antipolar off-centering.
Ge-like behavior persists upon alloying and gives way to Sn-like behavior only beyond roughly \qty{65}{\percent} Sn.
In the high-temperature phases, the B-site cations remain locally off-centered but directionally disordered.
On the Ge-rich side, the distorted inorganic framework biases the soft MA orientational landscape toward a restricted set of preferred directions.
This coupling shows how the composition of the inorganic sublattice can tune molecular ordering in lead-free hybrid perovskites.
\end{abstract}

\maketitle

\section{Introduction}

Halide perovskites, including the prototypical \ce{MAPbI3}, have emerged as one of the most promising classes of materials for photovoltaic applications, with lead-based perovskite solar cells now reaching power conversion efficiencies exceeding \qty{26}{\percent} \cite{nrel-efficiencies}.
Despite the remarkable progress, the toxicity of lead remains a concern, motivating the exploration of alternative B-site cations such as \ce{Sn^{2+}} and \ce{Ge^{2+}} \cite{Ke2019,Hao2014,Noel2014,Krishnamoorthy2015,Stoumpos2015}.
Sn-based compounds are currently the leading lead-free absorbers \cite{Hao2014,Noel2014,Ke2019,Mitzi1995}, and partial substitution of Sn by Ge has been reported to improve the oxidation resistance and device performance of several Sn-based compositions and film architectures \cite{Ito2018,Chen2019,Kama2022,Zhao2024}.
Beyond this practical appeal, the substitution of Pb by the lighter group-14 elements introduces qualitatively different structural chemistry that is of fundamental interest in its own right \cite{Radha2018,Li2022lonepair,HyltonFarrington2026}.

Mixing on the B-site is a well-established strategy for continuously tuning the structural, electronic, and optical properties of halide perovskites \cite{Dalpian2019,Im2015,Rimkus2025,Ju2017}.
Ge--Sn alloying is a particularly interesting case, since the two end members favor fundamentally different distortion mechanisms, which mixing brings into competition.
In \ce{MAGeI3}, the small ionic radius of \ce{Ge^{2+}} and its stereochemically active lone pair drive a pronounced polar off-centering of the B-site cation within the octahedral cage \cite{Stoumpos2015,Radha2018,Li2022lonepair}.
\ce{MASnI3}, by contrast, exhibits cooperative octahedral tilting instabilities, as does its Pb analog \cite{Stoumpos2013tin,Takahashi2011,Yang2017}, although weaker, antipolar off-centering tendencies of the Sn cation have been reported in the related compound \ce{CsSnBr3} \cite{Fabini2024,Fabini2016}.
These distortions have direct optoelectronic consequences: the photoluminescence intensity of \ce{MAGeI3} is maximized at an intermediate degree of off-centering, peaking under pressure at a bond-length distortion of $\mathcal{D} \approx 0.2$ compared with $\mathcal{D} = 0.32$ at ambient conditions \cite{Lu2021}, and the structural transitions of \ce{MASnI3} are accompanied by pronounced changes in carrier lifetime and defect-related recombination \cite{Parrott2016}.
The same lone-pair displacement makes the \ce{CsGeX3} (X = Cl, Br, I) perovskites ferroelectric \cite{Zhang2022,Kashikar2026}, and \ce{MAGeI3} is predicted to be comparably polar \cite{Zhao2017strong}.
The mixed \ce{MAGe_{1-x}Sn_xI3} system thus presents a setting in which polar off-centering, antipolar tendencies, octahedral tilting, and the molecular degrees of freedom coexist and compete.
A previous first-principles study examined the mechanical and optoelectronic properties of selected \ce{MAGe_{1-x}Sn_xI3} compositions in an assumed orthorhombic structure \cite{Hossain2020}, and samples spanning the full composition range have been characterized experimentally at room temperature by X-ray diffraction and optical absorption, with the optical gap increasing from \qty{1.3}{\electronvolt} at the Sn end to \qty{2.0}{\electronvolt} at the Ge end \cite{Nagane2018}.
The finite-temperature structural phase behavior of this system has, however, not been mapped.
In particular, little is known about how the organic MA sublattice responds to and interacts with the competing inorganic distortions.

The MA molecule carries a permanent dipole moment, and its orientational energy landscape is shaped by electrostatic, hydrogen-bonding, steric, and lattice-mediated interactions with the inorganic cage \cite{Li2016,Ding2023,Stroppa2015}.
In \ce{MAGeI3}, the strong $\langle 111\rangle$ Ge off-centering may therefore lift the near-degeneracy of several local MA orientational minima and favor a restricted subset of preferred crystallographic directions.
This behavior would contrast with \ce{MAPbI3}, in which the MA cations dynamically sample several symmetry-equivalent preferred orientations and reorient on picosecond timescales at ambient conditions \cite{Leguy2015,Mattoni2015,Chen2015,Lahnsteiner2016,Shimamura2016,Li2018,Shuck2019,fransson2023revealing,Laven2023}, and with \ce{MASnI3} and \ce{MASnBr3}, for which orientational disorder persists to comparatively low temperatures \cite{Swainson2010,Ha2020,Kubicki2020}.
Such polar molecular order would also be of interest for second-order nonlinear optics, since \ce{MAGeI3} exhibits a strong, phase-matchable second-harmonic-generation response \cite{Stoumpos2015}.
The nonlinear optical response is present in other Ge halide perovskites, both organic and inorganic, and has been shown to correlate with organic-cation ordering, chemical substitution, and framework distortion \cite{Ding2023,Tang2005,Lin2008,Qu2025}.
How the molecular orientational landscape evolves with composition and temperature, and whether molecular disorder feeds back on the stability of the inorganic polar order, remain unknown.

Here, we study the structural phase behavior of \ce{MAGe_{1-x}Sn_xI3} across the full composition range using large-scale \gls{md} simulations enabled by a \gls{mlip} \cite{FanZenZha21,fransson2023phase,Hainer2025}.
We find that the two end members order in fundamentally different ways: \ce{MAGeI3} develops strong long-range polar off-centering accompanied by MA orientational order with a much more gradual temperature dependence, whereas \ce{MASnI3} combines octahedral tilting with weaker, predominantly antipolar off-centering.
The Ge-like behavior proves remarkably persistent upon alloying, dominating the phase diagram up to a crossover at roughly \qty{65}{\percent} Sn.
Above the respective ordering transitions, the B-site cations of all compositions remain locally off-centered while losing their directional order, entering a dynamically disordered state that parallels the local off-centering observed by pair-distribution-function analysis in the cubic phases of Sn- and Pb-based perovskites \cite{Laurita2017,Fabini2016,Worhatch2008}.
Our results show that the MA orientations follow the B-site off-centering pattern across the full composition range and establish B-site mixing as a handle on the interplay of off-centering, tilting, and molecular order in this lead-free system.

\section{Results and discussion}

\subsection{Phase transitions}
\begin{figure*}[ht]
    \centering
    \includegraphics[scale=0.113]{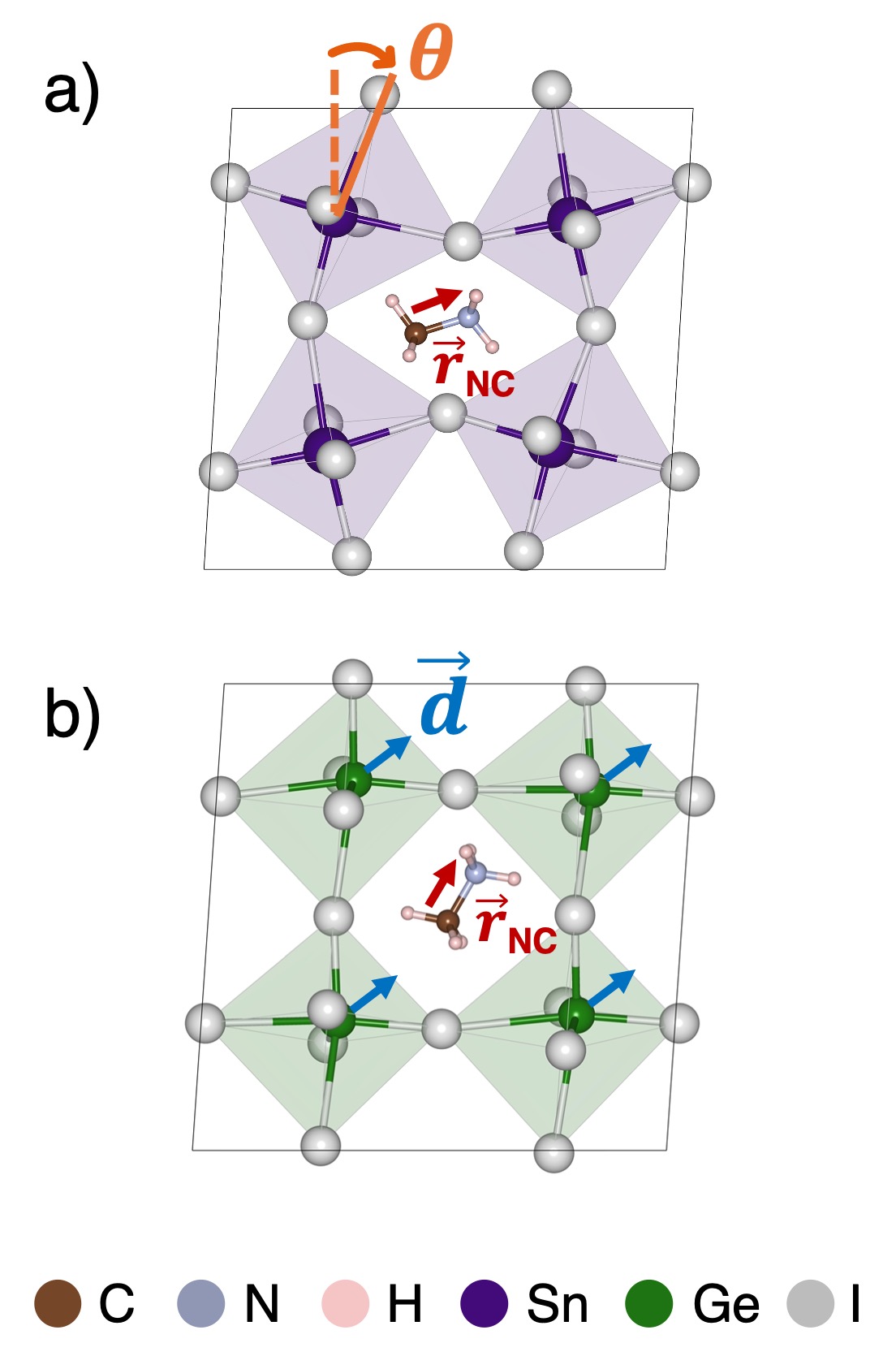}%
    \includegraphics[scale=0.96]{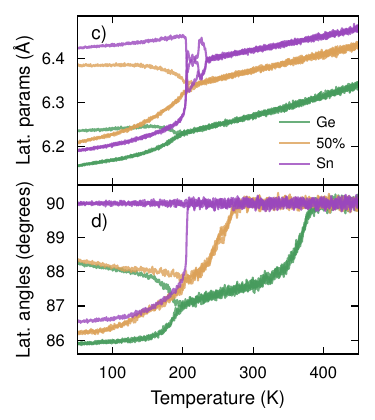}%
    \includegraphics[scale=0.96]{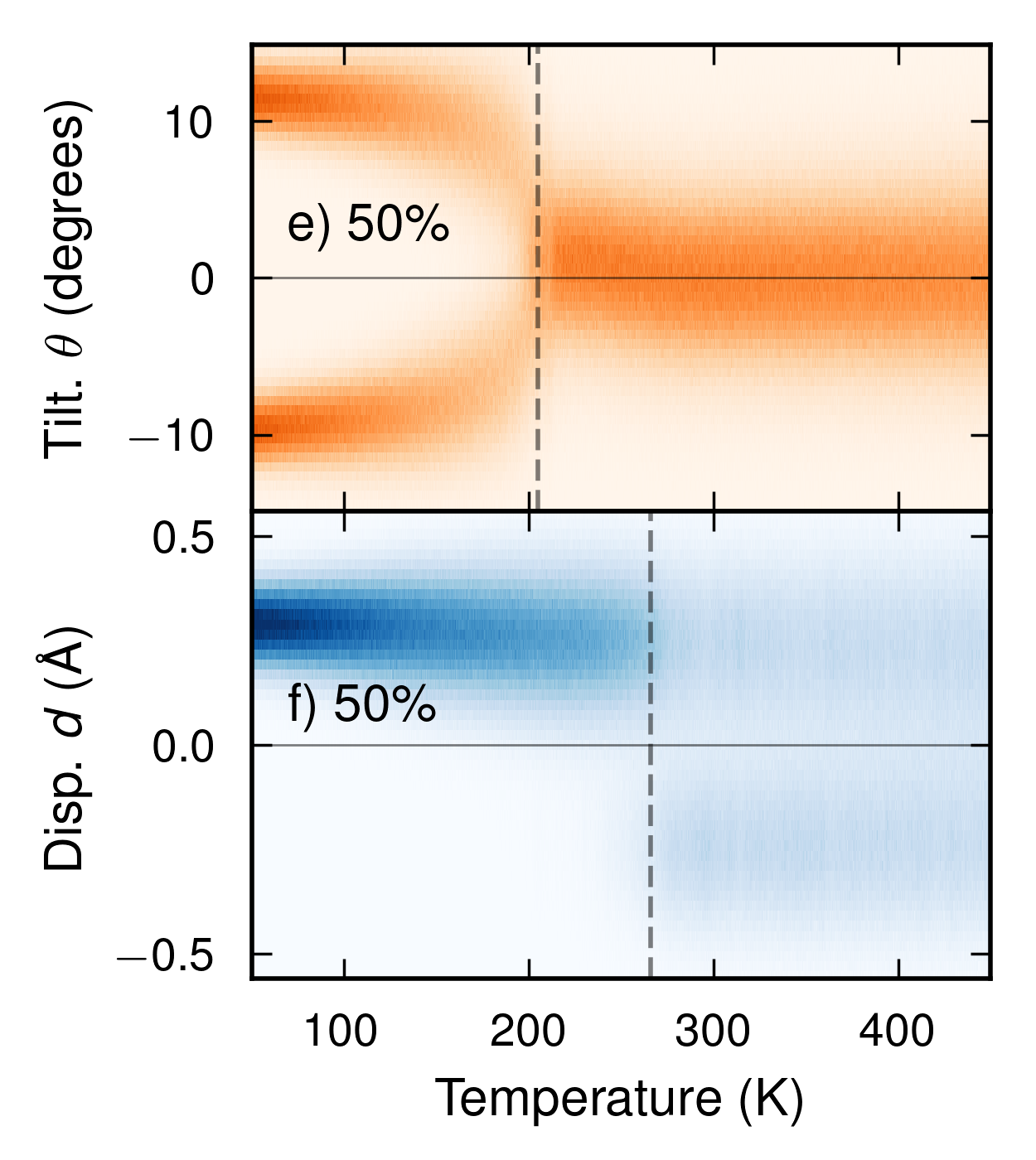}
    \caption{Atomistic structures of a) \ce{MASnI3} and b) \ce{MAGeI3}, where the octahedral tilt angle ($\theta$) and the off-centering displacement ($\vec{d}$) are indicated by orange and blue arrows, respectively.
    c) Lattice parameters and d) angles between the cell vectors along the heating runs of the Ge, the \qty{50}{\percent} mixed, and the Sn systems.
    The lattice parameters are computed as the lengths of the cell vectors of a supercell based on the conventional cubic cell, and the lattice angles correspond to the angles between the cell vectors.
    e, f) Distributions of e) the tilt angles ($\theta$) and f) the off-centering displacements ($d$), as indicated in a, b), along the heating run of the \qty{50}{\percent} mixed system.
    The vertical dashed lines indicate the two phase transitions of the \qty{50}{\percent} mixed system.
    }
    \label{fig:heating}
\end{figure*}

\begin{figure}[ht]
    \centering
    \includegraphics{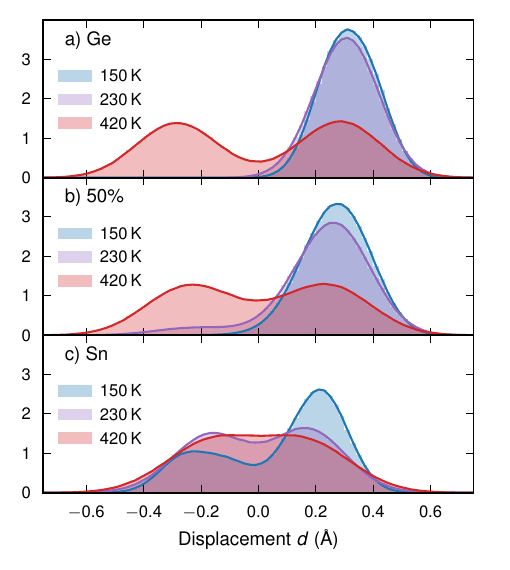}
    \caption{Distributions of the Cartesian components of the off-centering displacements $\vec{d}$, pooled over all B-site atoms, for a) Ge, b) the \qty{50}{\percent} mixture, and c) Sn at \qty{150}{\kelvin}, \qty{230}{\kelvin}, and \qty{420}{\kelvin}.
    }
    \label{fig:displacements_hists}
\end{figure}

\begin{figure}[ht]
    \centering
    \includegraphics[scale=0.11]{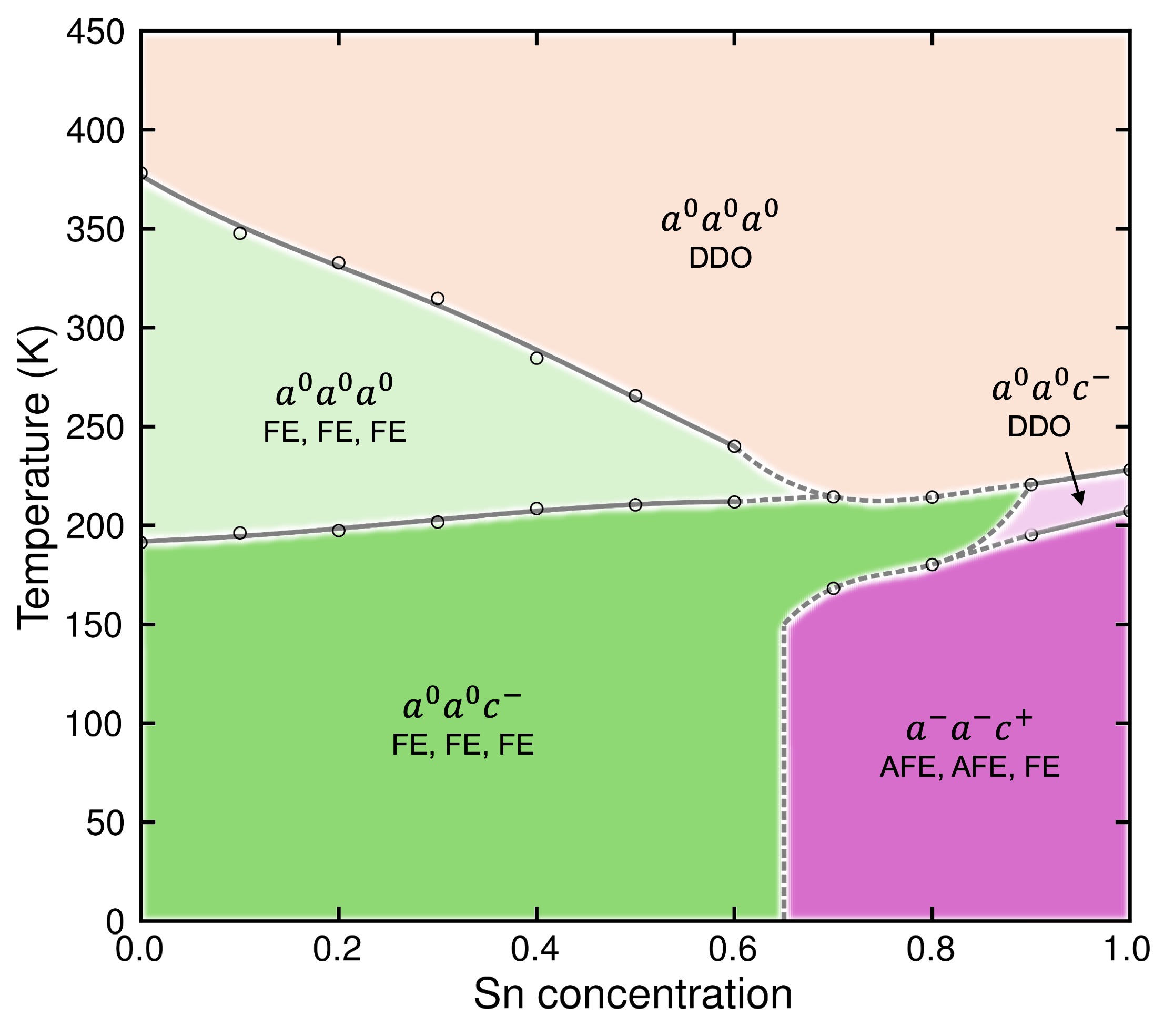}
    \caption{Predicted phase diagram of \ce{MAGe_{1-x}Sn_xI3}.
    Markers indicate points at which phase transitions are observed in the heating runs, and the lines serve as guides to the eye.
    Dashed lines correspond to regions of high uncertainty, where, for example, the exact phase was difficult to determine from the order parameters.
    The phases are labeled according to their off-centering order and octahedral tilt pattern.
    }
    \label{fig:phase_diagram}
\end{figure}

We determined the ground-state structures of the end members \ce{MAGeI3} and \ce{MASnI3} using a random structure search with the \gls{qnep}, following the approach previously applied to hybrid halide perovskites in Refs.~\cite{fransson2023revealing, Dutta2025}.
For the Ge system we find a structure exhibiting off-centering along the $\langle 111\rangle$ direction, combined with an out-of-phase tilt along one axis, as has also been found for \ce{CsGeBr3} \cite{Priyanka-paper}.
For the Sn system we find a tilt pattern resembling the previously predicted \hmn{Pnma} ground state, namely $a^-a^-c^+$ \cite{Ozorio2021}, but with additional Sn off-centering resulting in a lower symmetry.
The Sn off-centering is \gls{afe}, i.e., alternating between neighboring B sites, along two directions, and \gls{fe} along the third direction, which coincides with the $c^+$ tilt axis.
A closely related pattern has been reported for \ce{CsSnBr3}, where lone-pair-driven displacements are noncollinear within the plane normal to the unique octahedral tilt axis and ferroic along that axis \cite{Fabini2024}, in direct correspondence with the \gls{afe}/\gls{afe}/\gls{fe} decomposition found here.

Because the optimal starting configuration for the mixed compositions is not known a priori, we initialized every composition from both end-member ground states and compared the energies obtained along the corresponding heating runs (\autoref{sfig:U_vs_T}).
The \ce{MAGeI3} ground state yields the lower-energy starting point for Ge-rich compositions, whereas above roughly \qty{65}{\percent} Sn the \ce{MASnI3} ground state is preferred.
While this criterion is admittedly crude, it provides the best available estimate of the appropriate starting structure for each composition.

First, we consider the phase transitions observed in the heating runs for the end members and the \qty{50}{\percent} mixture (\autoref{fig:heating}).
For \ce{MAGeI3}, two clear transitions are observed in the lattice parameters and the angles between the cell vectors (\autoref{fig:heating}c,d), at \qty{200}{\kelvin} and around \qty{380}{\kelvin}.
These correspond to the system first losing its octahedral tilt (at \qty{200}{\kelvin}) and subsequently losing the long-range order of the B-site off-centering displacements (at \qty{380}{\kelvin}), as seen from the evolution of the order parameters (\autoref{sfig:global_order_parameters}).
For \ce{MASnI3} we also observe two transitions: a first one at \qty{200}{\kelvin}, at which the long-range order of the Sn off-centering breaks down and the tilt pattern changes to $a^0a^0c^-$, followed closely by a second transition at around \qty{230}{\kelvin}, at which the remaining tilt vanishes and the system adopts the high-symmetry cubic phase.
The \qty{50}{\percent} mixture follows the same transition sequence as the Ge end member, with transitions at \qty{200}{\kelvin} and \qty{280}{\kelvin}.
The corresponding structural changes can be identified from the distributions of the tilt angles and off-centering displacements (\autoref{fig:heating}e,f): the tilting vanishes at \qty{200}{\kelvin}, and the long-range order of the off-centering is lost at \qty{280}{\kelvin}.

A key difference between the Ge and Sn end members lies in the ordering of the B-site off-centering, with Ge exhibiting an \gls{fe} distortion.
The distributions of the Cartesian components of the displacements, pooled over all B-site atoms, are shown for three temperatures in \autoref{fig:displacements_hists}.
For Ge, all displacements are aligned up to the transition to the cubic phase, giving rise to a single peak at a finite displacement.
Above the transition, the distribution becomes a symmetric double Gaussian, showing that local off-centering persists while its long-range order is broken; direct visual inspection of the trajectories confirms repeated switching between the possible $\langle 111\rangle$ states, a behavior we refer to as \gls{ddo}.

The \qty{50}{\percent} mixture shows very similar behavior, albeit with slightly smaller displacements.
For Sn, the low-temperature distribution is asymmetric, reflecting the \gls{afe}/\gls{afe}/\gls{fe} pattern of the ground state: the two \gls{afe} components contribute symmetrically about zero, while the \gls{fe} component adds weight on one side only.
In the intermediate phase, the long-range order is lost and the distribution becomes a symmetric double Gaussian, i.e., the Sn sublattice likewise enters a \gls{ddo} state.
At \qty{420}{\kelvin}, the distribution appears unimodal and centered around zero, although local off-centering remains (\autoref{sfig:rdf}) and the distribution may equally well correspond to two strongly overlapping symmetric contributions.
In either case, this indicates that the local off-centering in Sn becomes increasingly dynamic with increasing temperature.
Altogether, this reflects the much stronger tendency toward off-centering in Ge than in Sn.

Experimentally, \ce{MASnI3} is reported to undergo two transitions, at about \qty{100}{\kelvin} and \qty{275}{\kelvin} \cite{Takahashi2011,Stoumpos2013tin}.
Our simulations likewise yield two transitions but compress them into a narrower window, at \qty{200}{\kelvin} and \qty{230}{\kelvin}.
We note that the reported ambient-temperature structure, \hmn{P4mm} \cite{Takahashi2011}, is polar rather than tilted, so that the experimental sequence may differ in character from the tilt-driven sequence found here.

The predicted structural parameters can be compared directly with experiment.
At \qty{300}{\kelvin}, \ce{MAGeI3} is tilt-free and ferroelectrically ordered in our simulations, corresponding to an \hmn{R3m}-like structure, in agreement with the rhombohedral structure reported experimentally at ambient conditions \cite{Stoumpos2015}.
We obtain a pseudocubic lattice parameter of \qty{6.27}{\angstrom} and a rhombohedral cell-angle deviation of \qty{2.4}{\degree}, close to the experimental values of \qty{6.18}{\angstrom} and \qty{2.5}{\degree} \cite{Stoumpos2015}.
The Ge--I first coordination shell is split into bond lengths of \qty{2.73}{\angstrom} and \qty{3.63}{\angstrom}, compared with experimental values of \qty{2.77}{\angstrom} and \qty{3.45}{\angstrom}, respectively \cite{Lu2021,Stoumpos2015}.
This corresponds to an off-centering distortion of $\mathcal{D} = \num{0.43}$ (\autoref{sfig:rdf}), compared with $\mathcal{D} = \num{0.32}$ from experiment \cite{Lu2021}.
Overall, the simulations reproduce the experimental symmetry, cell metric, and magnitude of the Ge off-centering distortion reasonably well.

For \ce{MASnI3}, the Sn--I first coordination shell at \qty{360}{\kelvin} is clearly split, with a bond-length separation of \qty{0.63}{\angstrom} and an off-centering distortion of $\mathcal{D} = \num{0.29}$ (\autoref{sfig:rdf}).
This is consistent with pair-distribution-function measurements at the same temperature, which show a strongly distorted local Sn--I environment associated with $\langle 111\rangle$ Sn off-centering, with a refined Sn displacement of approximately \qty{0.22}{\angstrom} \cite{Laurita2017}.
In contrast, applying the same analysis to \ce{MAPbI3} shows no resolved splitting of the Pb--I coordination shell, giving $\mathcal{D} = \num{0.0}$, consistent with the very weak or absent local Pb off-centering found experimentally \cite{Laurita2017}.
Together with \ce{MAGeI3}, these results reproduce the experimental trend from strong Ge off-centering, through substantial Sn off-centering, to very weak or absent off-centering in the Pb compound.

Lastly, we carry out order-parameter analysis, as described above, systematically across the composition range to map out the full phase diagram of the mixed system (\autoref{fig:phase_diagram}).
The diagram reveals a crossover between Ge-like and Sn-like behavior at around \qty{65}{\percent} Sn.
On the Ge-rich side, the transition temperature to the cubic phase decreases almost linearly from \qty{380}{\kelvin} to around \qty{230}{\kelvin} with increasing Sn content.
Near the crossover, compositions just above \qty{65}{\percent} Sn pass through a Ge-like \gls{fe} intermediate phase upon heating rather than the Sn-like $a^0a^0c^-$ phase.
We note that the transitions in this composition and temperature region are difficult to determine precisely (\autoref{sfig:tilt_disp_0p0}--\autoref{sfig:tilt_disp_1p0}).
Uncertainties stemming from the underlying \gls{dft} reference data, finite \gls{mlip} accuracy, and finite-rate and hysteresis effects mean that the transition temperatures and phase boundaries should be regarded as qualitative.
Moreover, the phase diagram describes the structural behavior of the sampled disordered alloys and does not account for possible chemical ordering, miscibility, or competing phases beyond those considered here.
What is clear, however, is the competition between the \gls{fe} off-centering on the Ge-rich side and the \gls{afe} tendencies and octahedral tilting on the Sn-rich side, which gives rise to the rich phase behavior of this mixed system.

\subsection{Molecular alignment}

\begin{figure}[ht]
    \centering
    \includegraphics[width=\linewidth]{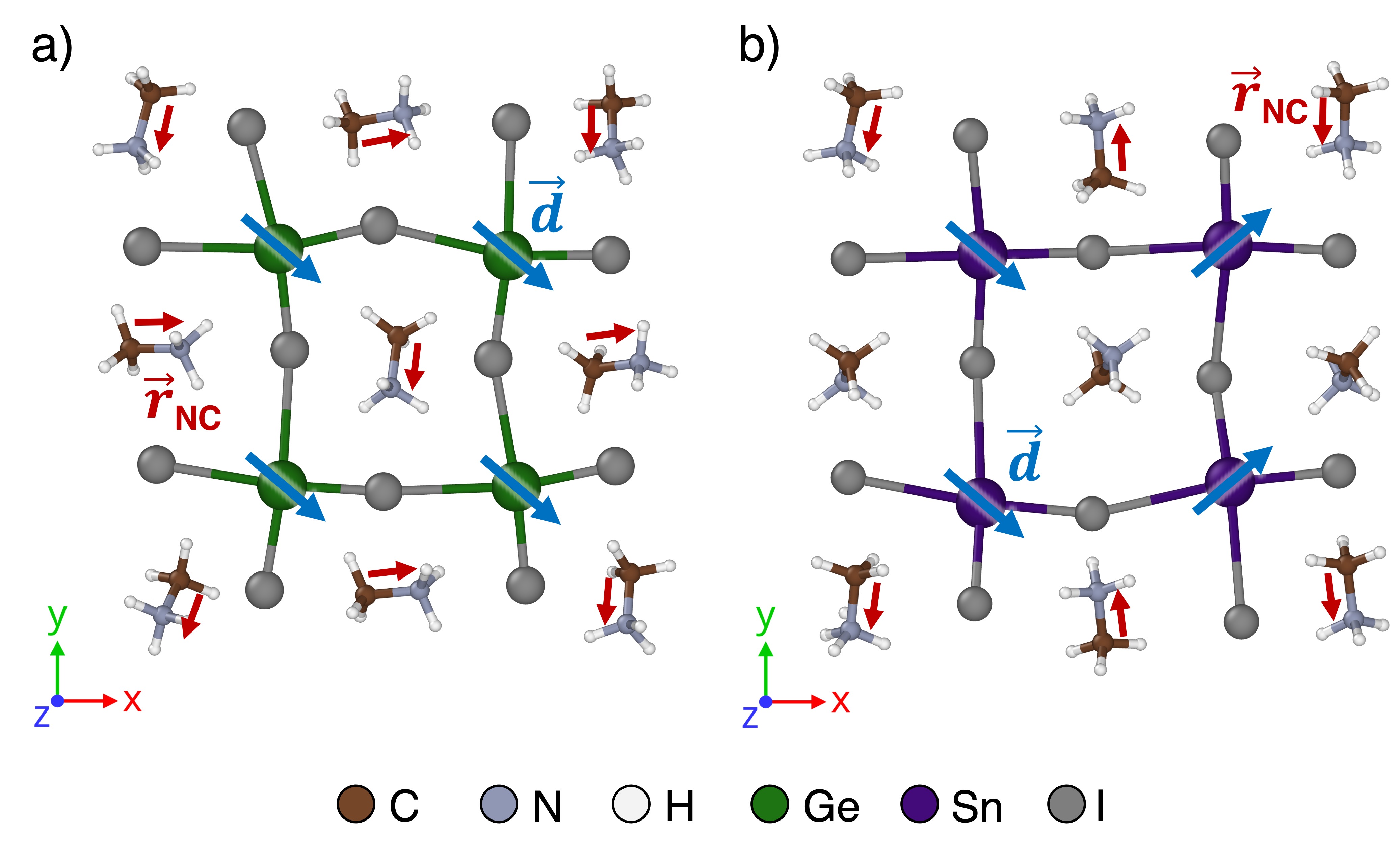}
    \caption{\Gls{md} snapshots at \qty{150}{\kelvin} for a) \ce{MAGeI3} and b) \ce{MASnI3}.
    Structures are rendered with \ovito{} \cite{Stukowski2010}, with the Ge/Sn bonds to neighboring iodine atoms shown.
    The blue arrows for the B-site indicate the off-centering displacements (relative to the neighboring iodine atoms), and the red arrows indicate the $\vec{r}_\mathrm{NC}$ vectors, pointing in the direction of the MA dipole.
    }
    \label{fig:snapshot_150K}
\end{figure}

\begin{figure*}[ht]
    \centering
    \includegraphics[scale=0.098]{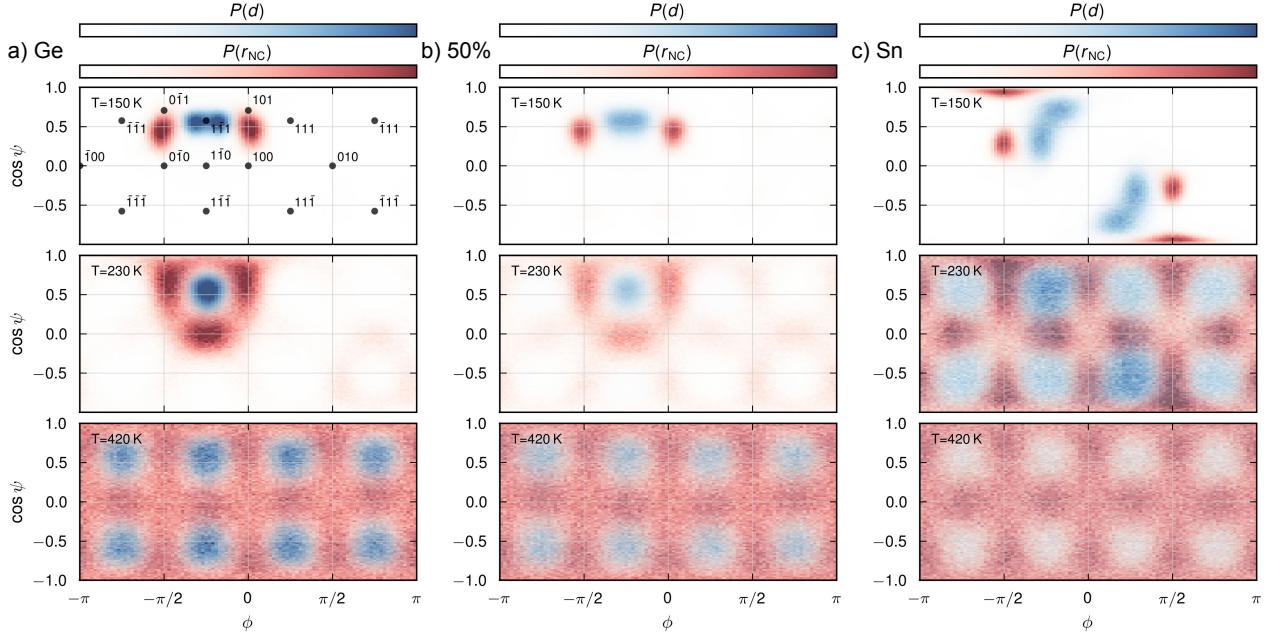}
    \caption{Distributions of the directions of the MA bond vectors ($\vec{r}_\mathrm{NC}$, red) and the B-site off-centering displacements ($\vec{d}$, blue) in a) Ge, b) the \qty{50}{\percent} mixture, and c) Sn, shown as a function of the spherical angles $\phi$ and $\psi$ at \qty{150}{\kelvin}, \qty{230}{\kelvin}, and \qty{420}{\kelvin}.
    The distributions are averaged over several snapshots corresponding to roughly $\pm$\qty{10}{\kelvin}.
    Blue maxima mark the directions of the B-site displacements and red maxima the directions of the MA bond vectors, so that adjacent blue and red maxima indicate alignment of the two.
    Each map covers the full sphere of directions; the eight blue maxima at \qty{420}{\kelvin} thus correspond to the eight $\langle 111\rangle$ directions of the dynamically disordered off-centering.
    Selected crystallographic directions are indicated in a), and the same mapping applies to all panels.
    The polar angle is plotted as $\cos\psi$ rather than $\psi$, so that equal areas on the map correspond to equal solid angles and an isotropic distribution of orientations appears uniform.
    Note that the color scales differ between temperatures in order to most clearly show the underlying distributions; in addition, the Sn panels in c) use different color scales at \qty{150}{\kelvin} and \qty{230}{\kelvin} than those in a, b).
    }
    \label{fig:MAd_heatmaps}
\end{figure*}

The B-site off-centering is accompanied by pronounced orientational ordering of the organic sublattice, particularly in the Ge-rich compositions.
The \ce{MA+} cation carries a formal charge of \qty{+1}{\elementarycharge} that is distributed asymmetrically between its two ends, with the ammonium group (\ce{NH3}) holding the larger share.
The molecule therefore carries a dipole moment directed from the methyl toward the ammonium group, i.e., along $\vec{r}_\mathrm{NC}$.
The relationship and strong correlation between the molecular orientations and the B-site displacement directions can be visualized directly in the structural snapshots from the \gls{md} simulations (\autoref{fig:snapshot_150K}).
For Ge (\autoref{fig:snapshot_150K}a), the MA cations clearly tend to point along the $\langle 110\rangle$ directions nearest to the $\langle 111\rangle$ Ge displacement direction, demonstrating a strong correlation between these two dipoles.
For Sn (\autoref{fig:snapshot_150K}b), the situation is more complex, since the Sn atoms partly adopt \gls{afe} distortions with alternating displacement directions, and the MA orientations follow the same alternating pattern.

The alignment of the B-site off-centering and the MA bond-vector orientation can be further quantified by considering the full distributions of the directions of $\vec{d}$ and $\vec{r}_\mathrm{NC}$ (\autoref{fig:MAd_heatmaps}).
The directions are represented by the spherical angles $\phi$ and $\psi$, where $\phi$ denotes the azimuthal angle in the $xy$ plane and $\psi$ the polar angle measured from the $z$ axis.
At \qty{150}{\kelvin}, all Ge atoms are off-centered along the same direction, corresponding to one of the eight $\langle 111\rangle$ directions, and the MA cations adopt only two preferred orientations, both of $\langle 110\rangle$ type and lying close to the off-centering direction.
This strong alignment persists to elevated temperatures.
At \qty{230}{\kelvin}, in the tilt-free intermediate phase, the MA cations explore a wider range of orientations and a third $\langle 110\rangle$-type orientation becomes populated, while the alignment with the Ge off-centering direction is retained.
This can be rationalized geometrically: for off-centering along, e.g., $[111]$, the three directions $[110]$, $[101]$, and $[011]$ all lie close to the off-centering axis, and the appearance of the third orientation is consistent with these directions becoming equivalent once the octahedral tilt is lost.
Upon the transition into the cubic phase, the Ge displacements enter the \gls{ddo} state, remaining locally off-centered while losing their long-range directional order, and consequently the MA cations lose their preferred orientations at the same transition.
The distinct thermal evolution of the B-site and MA global ordering is shown in \autoref{sfig:global_order_parameters}.
In \ce{MAGeI3}, the MA orientational order weakens progressively on heating, consistent with thermally activated reorientation, whereas the Ge polar order remains robust until sharply dropping to zero at the cubic transition.
The alignment reflects collective lattice ordering, with no notable additional nearest-neighbor angular correlation (\autoref{sfig:ma-bsite-angle}).

The \qty{50}{\percent} mixture shows the same qualitative trend as the Ge system, but the distributions of both the MA orientations and the B-site displacements are more diffuse.
This indicates that the preferred orientations are less strongly favored and that a larger degree of dynamic disorder is present in the mixed system.
For Sn, the picture is qualitatively different: the \gls{afe} pattern of the B-site displacements gives rise to two distinct displacement orientations at \qty{150}{\kelvin}.
This is accompanied by a bimodal distribution of the MA bond vectors, whose maxima largely coincide with the two Sn displacement directions, showing that the MA orientations align with the long-range B-site displacement pattern.
With increasing temperature, these narrow distributions broaden and become diffuse, and the system eventually becomes dynamically disordered with respect to both the molecular orientations and the off-centering.

Taken together, our results support a picture in which B-site off-centering reshapes the MA orientational landscape.
In Ge-rich compositions, the strong polar $\langle 111\rangle$ distortion favors a restricted set of MA orientations whose molecular dipoles align with the inorganic polar distortion, in contrast to the broader and dynamically disordered MA orientational landscape in tetragonal \ce{MAPbI3} \cite{Mattoni2015, Lahnsteiner2016, fransson2023revealing}.
Sn-rich compositions show a qualitatively different molecular ordering associated with the partially antipolar B-site displacement pattern.
B-site mixing can therefore provide a tunable handle for selecting preferred molecular orientations and coupled polar or antipolar ordering patterns in hybrid halide perovskites.

\section{Conclusions}
Combining a \gls{mlip} with large-scale \gls{md} simulations, we have mapped the structural phase behavior of the mixed hybrid halide perovskite \ce{MAGe_{1-x}Sn_xI3} across the full composition range.
The phase diagram is governed by the competition between \gls{fe} off-centering on the Ge-rich side and \gls{afe} off-centering tendencies combined with octahedral tilting on the Sn-rich side, with the crossover between the two regimes located at roughly \qty{65}{\percent} Sn.
In the high-temperature phases, the off-centering enters a \gls{ddo} state, in which the B-site cations remain locally displaced without long-range directional order.
We further find correlated long-range ordering of the MA orientations and the B-site displacement patterns, although their thermal evolution differs markedly.
On the Ge-rich side, the MA orientational order weakens progressively on heating while the long-range Ge displacement order remains robust until the cubic transition, consistent with a soft multi-minimum molecular landscape in which the ordered inorganic framework favors orientations aligned with the polar distortion.
The qualitatively different molecular ordering in Sn-rich compositions shows that B-site mixing can tune the MA orientational landscape and the resulting polar or antipolar ordering patterns.
Whether increasing molecular disorder contributes to the abrupt loss of Ge long-range order remains an open question.

\section{Methods}
\label{sect:methods}

\subsection{DFT}
\Gls{dft} calculations were carried out using the all-electron electronic-structure code \textsc{fhi-aims} \cite{blum2009ab}.
We employed the r$^2$SCAN50 hybrid functional \cite{Furness2020, Bursch2022}, which admixes \qty{50}{\percent} exact exchange, together with the \emph{intermediate} default basis set \cite{blum2009ab}.
No additional dispersion correction was applied.
This functional has been shown to be accurate for phase transitions in Ge perovskites \cite{Priyanka-paper, Kashikar2026}. 

Scalar relativistic effects were treated using the atomic zeroth-order regular approximation (atomic ZORA).
The electronic self-consistency cycle was converged to an electron-density threshold of \qty{e-6}{\elementarycharge\per\bohr\cubed}.

Brillouin-zone integrations were performed using automatically generated $k$-point meshes based on a $k$-point density parameter of 5.6, corresponding to an approximate reciprocal-space sampling of \qty{0.18}{\per\angstrom}.

\subsection{qNEP construction and training}
We constructed a \gls{qnep} \cite{FanZenZha21,Fan22,Fan2026qnep} using the \gpumd{} package (version 4.8) \cite{FanWanYin22,XuBuPan25}.
In this framework, each atomic partial charge is predicted by a neural network from the local descriptor vector, and electrostatic interactions are evaluated using \gls{qnep} mode 1, which includes real-space, reciprocal-space, and self-energy contributions.
We additionally included the screened nuclear repulsion potential of Ziegler, Biersack, and Littmark \cite{Ziegler1985} as a short-range baseline, as implemented in \gpumd{} \cite{Liu2023}.
The training set was built in an iterative way: starting from an initial set of structures, we repeatedly added snapshots drawn from \gls{md} simulations spanning a range of supercell sizes (up to \num{200} atoms) and temperatures, together with newly identified low-energy structures, and retrained the model at each iteration.
The initial candidate set combined reported \ce{MAGeI3} and \ce{MASnI3} structures \cite{Stoumpos2015,Stoumpos2013tin}, calculated Materials Project structures \texttt{mp-995236}, \texttt{mp-995238}, and \texttt{mp-1094059} \cite{Horton2025}, and \ce{MAPbI3}-derived prototypes in which Pb was replaced by Ge or Sn.
The final potential was trained on \num{668} structures.
The training (test) root-mean-square errors are \qty{4.33}{\milli\electronvolt\per\atom} (\qty{7.67}{\milli\electronvolt\per\atom}) for energies, \qty{126}{\milli\electronvolt\per\angstrom} (\qty{188}{\milli\electronvolt\per\angstrom}) for forces, \qty{19.9}{\milli\electronvolt\per\atom} (\qty{18.5}{\milli\electronvolt\per\atom}) for virials, and \qty{0.0716}{\elementarycharge} (\qty{0.0596}{\elementarycharge}) for Born effective charges.
The corresponding parity plots are shown in \autoref{sfig:parity_plots}.

\subsection{Molecular dynamics}
All structure handling and the interface to \gpumd{} were managed through \ase{} \cite{Larsen2017} and \calorine{} \cite{calorine}.
Production \gls{md} simulations were performed with \gpumd{} \cite{FanWanYin22,XuBuPan25} using a timestep of \qty{0.5}{\femto\second} and supercells containing \num{49152} atoms.
In all mixed supercells, Ge and Sn were assigned randomly to the B-site sublattice, and the resulting occupational configuration remained fixed throughout each \gls{md} trajectory.

All simulations were run in the $NPT$ ensemble using stochastic cell rescaling \cite{Bernetti2020}, as implemented in \gpumd{}.
Phase transitions are observed through heating runs, which are ramped in temperature continuously from \qty{0}{\kelvin} to \qty{450}{\kelvin} over \qty{20}{\nano\second}, see \autoref{sfig:convergence} for convergence testing.

We characterized the local and global structure across composition and temperature through three quantities, which serve as order parameters for identifying the structural phases of the system (see \autoref{fig:heating}a,b).
Local octahedral tilt angles, $\theta_i^\alpha$, for B-site atom $i$ in direction $\alpha$ were computed following the procedure of Ref.~\cite{Wiktor2023}.
The orientation of each MA cation was tracked through its C--N bond vector, $\vec{r}_\mathrm{NC} = \vec{r}_\mathrm{N} - \vec{r}_\mathrm{C}$.
The B-site off-centering was quantified by the displacement
\begin{align}
    \vec{d} = \vec{r}_\mathrm{B} - \frac{1}{6}\sum_{i=1}^{6} \vec{r}_{\mathrm{I},i},
\end{align}
where $\vec{r}_\mathrm{B}$ is the position of the Ge or Sn atom and the second term is the geometric center of the surrounding octahedron, defined by its six neighboring iodine atoms.
A related distortion parameter that measures the B-site off-centering is the off-centering distortion $\mathcal{D}$ introduced in Ref.~\cite{Lu2021}, defined as
\begin{align}
    \mathcal{D} = \sum_{\alpha=1}^{3} \frac{|a_\alpha - b_\alpha|}{a_\alpha + b_\alpha},
\end{align}
where $a_\alpha$ and $b_\alpha$ are the short and long B--I bond lengths along the three trans I--B--I axes of the \ce{BI6} octahedron.
Unlike $\vec{d}$, $\mathcal{D}$ is dimensionless and directly accessible from diffraction, which enables a comparison with experimental values.

Octahedral tilting and $\langle 111\rangle$ off-centering transform under different irreducible representations and cannot hybridize directly, although they compete energetically \cite{Radha2018,HyltonFarrington2026}.
We therefore track them as separate order parameters.
Here, the labels \gls{fe} and \gls{afe} describe ferroic and antiferroic spatial patterns of B-site off-centering, respectively.
The phases and transitions were identified using these order parameters from heating runs across the composition range, together with the lattice parameters.
Additionally, these order parameters, which are local in nature, were also used to get a better understanding of the local structure of the systems.

\section*{Supporting information}
Parity plots for the \gls{qnep}; convergence tests with respect to system size, heating rate and independent trajectories; global B-site and MA order parameters; the angle between the B-site off-centering and the MA orientation; potential energies along heating and cooling runs; and composition-resolved distributions of tilt angles and off-centering displacements for $x = 0.0$ to $1.0$.

\section*{Data availability}
The \gls{qnep} model and reference \gls{dft} data generated in this study are openly available via Zenodo at \url{https://doi.org/10.5281/zenodo.21863901}.

\begin{acknowledgments}
Funding from the Swedish Foundation for Strategic Research through the Future Research Leader programme (FFL21-0129), the Swedish Energy Agency (grant No. 45410-1), the Swedish Research Council (grant Nos. 2018-06482, 2019-03993, 2020-04935, 2025-03999, and 2025-04685), the European Research Council (ERC Starting Grant No. 101162195), the Knut and Alice Wallenberg Foundation (grant Nos. 2023.0032 and 2024.0042), and the Area of Advance Nano at Chalmers is gratefully acknowledged. The computations were enabled by resources provided by the National Academic Infrastructure for Supercomputing in Sweden (NAISS) at C3SE, PDC, and NSC, partially funded by the Swedish Research Council through grant agreement No. 2022-06725, as well as by the Berzelius resource provided by the Knut and Alice Wallenberg Foundation at NSC.

The authors used AI tools (Claude, Anthropic, and ChatGPT, OpenAI) during the preparation of this manuscript for assistance with writing and language improvements.
All scientific content, interpretation, and final wording were reviewed and approved by the authors.
\end{acknowledgments}

\section*{Competing interests}
The authors declare no competing financial interest.

\bibliography{references}

\end{document}